\documentstyle[preprint,aps,epsfig]{revtex}

\def\be{\begin{equation}}
\def\ee{\end{equation}}
\def\ep{\epsilon}

\def\bea{\begin{eqnarray}}
\def\eea{\end{eqnarray}}

\begin{document}
\draft

\title{Computer simulations of polydisperse ER fluids in DID model}
\author{Andrew C. T. Wong$^1$, Hua Sun$^{1,2}$ and K. W. Yu$^1$}
\address{$^1$Department of Physics, The Chinese University of Hong Kong,
 Shatin, NT, Hong Kong \\
 $^2$Department of Physics, Suzhou University, Suzhou 215006, China}
\maketitle

\begin{abstract}
The theoretical investigations on electrorheological (ER) fluids are usually 
concentrated on monodisperse systems. 
Real ER fluids must be polydisperse in nature, i.e., 
the suspended particles can have various sizes and/or different 
dielectric constants.
An initial approach for these studies would be the point-dipole (PD) 
approximation, which is known to err considerably when the particles 
approach and finally touch due to multipolar interactions. 
In a recent work, we proposed a dipole-induced-dipole (DID) model for 
computer simulation of ER fluids, which was shown to be both more accurate 
than the PD model and easy to use. 
The DID model was applied to simulate the athermal aggregation of
particles in ER fluids and the aggregation time was found to be 
significantly reduced as compared to the PD model.
In this work, we will report results for the case when the dielectric 
contrasts of some particles can be negative. In which case, 
the direction of the force is reversed. Moreover, the inclusion of DID 
force further complicates the results because the symmetry between 
positive and negative contrasts will be broken by the presence of 
dipole-induced interactions. 
\end{abstract}
\vskip 5mm 
\pacs{PACS Number(s): 83.80.Gv, 82.70.-y, 42.20.-q}

\section{Introduction}

Many theoretical investigations on electrorheological (ER) fluids are 
usually concentrated on monodisperse systems in which all the suspended 
particles are of the same size and dielectric constant.
In reality ER fluids must be polydisperse in nature, i.e., 
the particles can have various sizes and/or different dielectric 
permittivities \cite{Ahn}.
For instance, the particle size has a significant impact on the yield
stress \cite{Ota}, as well on the rheology \cite{Lemaire}.
An initial approach for these studies would be the point-dipole (PD) 
approximation \cite{K1}. 
As many-body and multipolar interactions between the particles have been 
neglected, the PD approximation is known to err considerably when the 
particles approach and finally touch. 
The PD approximation becomes even worse when the dielectric contrasts 
between the suspended particles and the host medium become large. 
To circumvent the problem, we recently employed the multiple image method 
to compute the interparticle force for a polydisperse ER fluid \cite{MID}.
From the results, we proposed a dipole-induced-dipole (DID) model for 
computer simulations of ER fluids; the DID model yields very good 
agreements with the multiple image results for a wide range of dielectric 
contrasts and polydispersity \cite{MID}.
The DID model has recently been employed to simulate the athermal 
aggregation of particles in ER fluids in which the particles are of 
different permittivities \cite{Siu}. 
Moreover, the dielectric contrasts between the particles and the host 
fluids were all positive. 
The aggregation time was found to be significantly reduced,
both in uniaxial and rotating electric fields \cite{Siu}.
In this work, we will report results for the case when the 
dielectric contrasts of some particles can be negative. In which case, 
the direction of the force should be reversed \cite{Kaski}. 
The inclusion of the DID force further complicates the results because 
the symmetry between positive and negative contrasts will be broken by 
the presence of dipole-induced interactions. We found that the 
aggregation time can be much increased as compared to the PD model.

In the next section, we review the multiple image method and establish 
the DID model. In section III, we apply the DID model to the 
computer simulation of ER fluids in a uniaxial field. 
In section IV, we extend the simulation to athermal aggregation in rotating 
fields. Discussion on our results will be given. 

\section{multiple image method}

In this section, we generalize the multiple image method \cite{MID} to 
handle both positive and negative dielectric contrasts.
Consider a pair of dielectric spheres, of radii $a$ and $b$, dielectric 
constants $\ep_1$ and $\ep_1'$ respectively, separated by a distance $r$. 
The spheres are embedded in a host medium of a dielectric constant $\ep_2$. 
Upon the application of an electric field $\bf E_0$, the induced-dipole 
moment inside the spheres are, respectively, given by (SI units)
\be
p_{a0}=4 \pi \ep_0 \ep_2 \beta E_0 a^3,\;\;\; 
p_{b0}=4 \pi \ep_0 \ep_2 \beta' E_0 b^3,
\ee
where the dipole factors $\beta$, $\beta$' are defined as
\be
\beta=\frac{\ep_1-\ep_2}{\ep_1+2\ep_2},\;\;\; 
\beta'=\frac{\ep_1'-\ep_2}{\ep_1'+2\ep_2}.
\ee
In the point-dipole (PD) model, the force between two particles is given by
\be
{\bf F}_{PD} = {F \over r^4}
 \left[(2\cos^2 \theta - \sin^2 \theta)\hat{\bf r} 
 + \sin 2\theta \hat{\theta}\right],
\ee
where $F=12 \pi \ep_0 \ep_2 \beta \beta' a^3 b^3 E_0^2$, $\hat{\bf r}$ and
$\hat{\theta}$ are unit vectors.
Many previous studies were concentrated on the case in which both 
$\beta, \beta' > 0$. 
When the $\beta$s adopt opposite signs, the direction of the PD force 
will be reversed. In a recent paper, we derived a correction to the PD 
force from the multiple image method \cite{Siu}. 
The total dipole moment inside sphere $a$ is 
\bea
p_{aT}&=&(\sinh \alpha)^3 \sum_{n=1}^{\infty}
\left[\frac{p_{a0} b^3(-\beta)^{n-1}(-\beta')^{n-1}}
 {(b \sinh n \alpha +a \sinh(n-1) \alpha )^3}
+\frac{p_{b0} a^3(-\beta)^n(-\beta')^{n-1}}{(r \sinh n \alpha)^3}\right] ,
\label{mideqt1}
\\
p_{aL}&=&(\sinh \alpha)^3 \sum_{n=1}^{\infty}
\left[\frac{p_{a0} b^3(2\beta)^{n-1}(2\beta')^{n-1}}
{(b \sinh n \alpha +a \sinh(n-1) \alpha )^3}
+\frac{p_{b0} a^3(2\beta)^n(2\beta')^{n-1}}{(r \sinh n \alpha)^3}\right] ,
\label{mideqt2}
\eea
where the subscripts $T(L)$ denote a transverse (longitudinal) field, 
i.e., the applied field is perpendicular (parallel) to the line joining 
the centers of the spheres. 
Similar expressions for the total dipole moment inside sphere $b$ can be 
obtained by interchanging $a$ and $b$, as well as $\beta$ and $\beta'$. 
The parameter $\alpha$ satisfies 
\be
\cosh \alpha=\frac{r^2-a^2-b^2}{2 a b}.
\ee
The forces between the spheres is given by \cite{jackson}
\be
F_T=\frac{E_0}{2}\frac{\partial}{\partial r}(p_{aT}+p_{bT}), \;\;\; 
F_L=\frac{E_0}{2}\frac{\partial}{\partial r}(p_{aL}+p_{bL}).
\ee
It should be noted that the multiple image results can be used to compare 
among the various models according to how many terms are retained in the 
multiple image expressions: (a) PD model: $n=1$ term only, 
(b) DID model: $n=1$ to $n=2$ terms only, and 
(c) multiple-induced-dipole (MID) model: $n=1$ to $n=\infty$ terms.

For convenience, we define the reduced separation $\sigma=r/(a+b)$. 
Here we set $a=b$. 
We consider two cases: (a) $\beta>0, \beta'<0$ or $\ep_1>\ep_2>\ep_1'$ 
and (b) $\beta=\beta'<0$ or $\ep_2>\ep_1=\ep_1'$. 
The interparticle forces in the longitudinal and transverse cases 
are plotted in Fig.\ref{figforce} with different values of the dielectric 
contrasts. 
For the case (a) depicted in Fig.\ref{figforce}(a), 
the magnitude of MID force falls between that of the PD and DID in both 
the longitudinal and transverse cases. 
At low contrasts, the DID results almost coincide with the MID results. 
While at high contrast, the DID model exhibits significant deviation from 
MID when $\sigma<1.1$. 
For the case (b) as depicted in Fig.\ref{figforce}(b), the DID results 
almost coincide with the MID results at low contrast, while deviate 
significantly at high contrast.
The forces are found to be qualitatively different from the 
case of $\beta, \beta' > 0$. 
Thus, by including the multiple image contributions, it is observed that the 
case of $\beta, \beta'<0$ significantly different from the presumably 
symmetric case of $\beta, \beta'>0$. The symmetry has been broken due to 
the presence of DID forces.

\section{athermal aggregation in the uniaxial field}

The multiple image expressions [Eqs.(\ref{mideqt1}) and 
(\ref{mideqt2})] allow us to calculate the 
correction factor defined as the ratio between the DID and the PD forces 
\cite{Siu}:
\bea
\frac{F_{DID}^{(\parallel)}}{F_{PD}^{(\parallel)}}
&=&1+\frac{2\beta a^3 r^5}{(r^2-b^2)^4}
+\frac{2\beta' b^3 r^5}{(r^2-a^2)^4}
+\frac{4\beta \beta' a^3 b^3 (3r^2-a^2-b^2)}{(r^2-a^2-b^2)^4},
\label{K-para} 
\\
\frac{F_{DID}^{(\perp)}}{F_{PD}^{(\perp)}}
&=&1-\frac{\beta a^3 r^5}{(r^2-b^2)^4}
-\frac{\beta' b^3 r^5}{(r^2-a^2)^4}
+\frac{\beta \beta' a^3 b^3 (3r^2-a^2-b^2)}{(r^2-a^2-b^2)^4}, 
\\
\frac{F_{DID}^{(\Gamma)}}{F_{PD}^{(\Gamma)}}
&=&1+\frac{\beta a^3 r^3}{2 (r^2-b^2)^3}
+\frac{\beta' b^3 r^3}{2(r^2-a^2)^3}
+\frac{3\beta \beta' a^3 b^3 }{(r^2-a^2-b^2)^3},
\label{K-Gamma}
\eea
where $F_{PD}^{(\perp)}={3 p_{a0}p_{b0}}/{4 \pi \ep_0 \ep_2 r^4}$, 
$F_{PD}^{(\parallel)}={-6 p_{a0}p_{b0}}/{4 \pi \ep_0 \ep_2 r^4}$, and 
$F_{PD}^{(\perp)}={-3 p_{a0}p_{b0}}/{4 \pi \ep_0 \ep_2 r^4}$ 
are the point-dipole forces for the transverse, longitudinal, and $\Gamma$ 
cases, respectively.
If we denote the ratios in Eqs.(\ref{K-para})--(\ref{K-Gamma}) 
by $K_\parallel, K_\perp$ and $K_\Gamma$ respectively, 
the force between two particles is modified to
$$
{\bf F}_{DID} = {F \over r^4}
 \left[(2K_\parallel\cos^2 \theta - K_\perp\sin^2 \theta)\hat{\bf r} 
 + K_\Gamma\sin 2\theta \hat{\theta}\right].
$$

In what follows, we consider two spheres of equal radius $a$ and various 
dipole factors $\beta$ and $\beta'$ respectively. These spheres are 
initially at rest and at a separation $d_0$. For aggregation induced by 
a uniaxial field, we consider two cases. For case (a) $\beta>0, 
\beta'<0$ or $\ep_1>\ep_2>\ep_1'$, the electric field is perpendicular to 
the line joining the centers of the spheres. The equation of motion is 
given by 
\be
\frac{dz}{dt}=F_{\perp}(2z),
\label{eom1}
\ee
while for case (b) $\beta=\beta'<0$ or $\ep_1=\ep_1'<\ep_2$, 
the electric field is parallel to the line joining the centers of 
the spheres. The equation of motion becomes
\be
\frac{dz}{dt}=F_{\parallel}(2z),
\label{eom2}
\ee
where $z$ is the displacement of one sphere from the center of mass. 
The separation between the two spheres is therefore $d=2 z$ and the 
initial condition is $d=d_0$ at $t=0$. Eqs.(\ref{eom1}) and (\ref{eom2}) 
are dimensionless equation. 
Following Klingenberg and with slight modification, 
we choose the following natural scaling units to define 
the dimensionless variables \cite{K1}: 
\be
z_0=2a,\;\;\; t_0=12 \pi \eta_c a^2/F_0, \;\;\; F_0=\frac{3}{4} \pi 
\ep_0 \ep_2 E_0^2 a^2 ,
\ee
where $E_0$ is the field strength, $m$ the mass, and $\eta_c$ the 
coefficient of viscosity. 
We have followed Klingenberg to ignore the inertial effect and thermal 
motion of the particles \cite{K1}. 
The initial separation $d_0$ is related to the volume fraction $\phi$ by
\be
\frac{d_0}{2a}=\left(\frac{\pi}{6 \phi}\right)^{1/3}.
\ee
In the PD approximation, Eq.(\ref{eom1}) admits an analytic solution
\be
z=\left[\left(\frac{d_0}{4a}\right)^5+ \frac{5\beta\beta't}{16}\right]^{1/5},
\label{PD1}
\ee
while Eq.(\ref{eom2}) gives
\be
z=\left[\left(\frac{d_0}{4a}\right)^5- \frac{5\beta\beta't}{8}\right]^{1/5}.
\label{PD2}
\ee
It is obvious the above equations impose a condition for  aggregation in 
uniaxial field, namely, the transverse case requires that $\beta$ and 
$\beta'$ are of different signs while the longitudinal case requires that 
they are of the same sign.

For the DID model, we integrate the equations of motion by the 
fourth-order Runge-Kutta algorithm, with the time step $\delta t=0.0001$ 
for both small and large volume fractions. In Fig.\ref{figagguf}, 
we plot the reduced separation $\sigma$ against dimensionless time $t/t_0$. 
The results reveal that in general the DID results deviate slightly from 
the PD results at low volume fractions, i.e. at a large initial 
separation, while the deviation becomes large at high volume fractions. 
The deviations are more pronounced at high contrasts, attributed to a 
large attractive force, resulting in a smaller aggregation time. 
In both cases (a) and (b), the DID aggregation time is generally larger 
than that of PD. While in the case $\beta, \beta'>0$, the situation is 
reversed. This feature can be understood from the force magnitudes of the 
two models (see Fig.\ref{figforce}).

Fig.\ref{figratiouf} shows the ratio of the aggregation time for the PD 
model to the DID model against reduced initial separation $d_0/2a$. 
The aggregation time of the DID model is generally larger than that of 
the PD model especially when the initial separation is small. 
The correction factor is more pronounced at high contrast. 
In the case $\beta=\beta'<0$ or $\ep_2>\ep_1=\ep_1'$, we observe a 
non-monotonic dependence, the correction factor increases significantly 
when the spheres are close and decrease again when $d/2a=1.13$ or less at 
high contrast. This behavior can be understood from the interparticle 
force (Fig.\ref{figforce}(b), the three panels on the right). 
At low contrast, the difference in magnitude between the PD and DID force 
increases monotonically as the separation decreases. 
However at high contrast the difference increases first then decreases 
again as the separation decreases.

\section{athermal aggregation in the rotating field}

Perhaps it is more interesting to consider aggregation in a rotating 
electric field \cite{Siu}. Consider a rotating field applied in the 
$x$-$y$ plane, $E_x=E_0 \cos{\omega t}$, $E_y=E_0 \sin{\omega t}$. 
The dimensionless equation of motion for the two spheres becomes 
\be
\frac{dx}{dt}=F_{\parallel}(r) \cos^2{\omega t} + F_{\perp}(r) 
\sin^2{\omega t}, \;\;\; 
\frac{dy}{dt}=-F_{\Gamma}(r) \sin{2\omega t}
\ee
where $(x, y)$ are the displacement of one sphere from their center of 
mass and $r=2\sqrt{x^2+y^2}$ is the separation between two 
particles. In case of a large $\omega$, we may safely neglect the $y$ 
direction of the motion. For the PD approximation, the dimensionless forces 
are $F_{\parallel}=-2\beta\beta'/r^4$ and $F_{\perp}=\beta\beta'/r^4$, 
respectively, which yields the analytic solution 
\be
x=\left[\left(\frac{d_0}{4a}\right)^5
 -\frac{5\beta\beta'}{64\omega}(\omega t + 3 \sin{2 \omega t})\right]^{1/5} 
\ee
The separation between two spheres is $d=2x$. For the DID model, 
we integrate the equation of motion by the fourth-order Runge-Kutta 
algorithm, with a time step $\delta t=1/400\omega$. 

Note that the two particles cannot aggregate if $\beta>0, \beta'<0$ or 
$\ep_1>\ep_2>\ep_1'$ in the rotating field because the repulsive force 
induced by longitudinal field $F_L$ is always larger than the attractive 
force induced by transverse field $F_T$, while the particles spend equal 
times in both fields on the average. The displacement-time graph of the 
aggregations with $\ep_2>\ep_1=\ep_1'$ in the two models are plotted in 
Fig.\ref{figaggrf}. The shape of the displacement-time graph of DID 
model is different from that of the PD model, especially when the two 
spheres are close, resulting in a large deviation between the DID and PD 
models in the aggregation time. This behavior can again be understood by 
the interparticle force between the two spheres. We can see from 
Fig.\ref{figforce}(b) that the DID longitudinal attractive force and its 
transverse repulsive force is about the same when the separation is 
small. Since the spheres spend equal times in the longitudinal and 
transverse field and they change the direction of motion according to the 
rotating field, the displacement of the spheres can be estimated by the 
difference between the magnitude of the longitudinal attractive force and 
the transverse repulsive force, as there would be no displacement if 
the magnitude of the two forces are the same. Thus when the two spheres 
are near, their velocity become smaller and have different shape in the 
displacement-time graph. On the other hand, the PD attractive force is 
sufficiently larger than its repulsive force at any separation.

We further calculated the correction factor of the aggregation time for 
the PD model with respect to the DID model in the same way as in the 
uniaxial field case. 
We plot the correction factor against the reduced initial separation 
$d_0/2a$ with different parameters in Fig.\ref{figratiorf}. 
The aggregation time is significantly increased when the mutual 
polarization of the two spheres is taken into account. The increase becomes 
even more pronounced and oscillating when the initial separation is 
small. The oscillation is due to the sensitive dependence on the initial 
polarization of dipoles when the spheres are close.

\section*{Discussion and Conclusion}

Here a few comments on our results are in order.
In this work, we have concentrated on the multipolar interactions between
touching particles, which we believe are more important than the many-body
or local-field effects \cite{Davis,Wang}. In the latter approaches, 
the particles in ER fluids are still treated as point dipoles, 
while their dipole moments are determined by adding the local-field 
corrections. 
In the DID model, the additional terms arise from multipole interactions, 
rather than from local-field corrections.

We have studied the aggregation time for two spherical particles. 
We should also examine the morphology of aggregation in polydisperse 
ER fluids, due to the dipole-induced forces. 
To this end, it may be difficult to obtain the ground state lattice 
in polydisperse ER fluids by using dynamic simulations alone, 
since both body-centred tetragonal and simple cubic or even honeycomb 
stacking lattices coexist in such systems according to Kaski and coworkers 
\cite{Kaski2}. Moreover, many external factors such as the numbers of 
particles, the size of the cell, etc. will affect the final results. 
We are currently trying to incorporate the DID terms in the interaction 
energy. In this connection, we can also examine the recently proposed 
structural transformation by applying the uniaxial and rotating fields 
simultaneously to an ER fluid \cite{Lo}.

In summary, we have used the DID model to deal with computer simulations
of polydisperse ER fluids for the case when the dielectric contrasts of 
some particles can be negative. We studied athermal aggregation of 
two spherical particles both in uniaxial and rotating electric fields.
We showed that an inclusion of the DID  force breaks the symmetry 
between positive and negative contrasts. 
As a result, the aggregation time can be much increased as compared to 
the PD model.

\section*{Acknowledgments}
This work was supported in part by the Direct Grant for Research, 
and in part by the RGC Earmarked Grant.
K. W. Y. acknowledges useful discussion with Dr. Jones T. K. Wan of the 
Princeton University.

\begin{figure}[h]
\caption{The interparticle force of the PD and DID models plotted against 
the reduced separation $\sigma$ between two spherical particles for 
several dielectric contrasts, $\ep_1/\ep_2$ for one particle and 
$\ep_1'/\ep_2$ for the other, both in the transverse ($T$) and 
longitudinal ($L$) electric fields: (a) $\beta>0, \beta'<0$ or 
$\ep_1>\ep_2>\ep_1'$ and (b) $\beta=\beta'<0$ or $\ep_2>\ep_1=\ep_1'$.} 
\label{figforce} 
\end{figure}

\begin{figure}[h]
\caption{The displacement-time graph for athermal aggregation of two 
spherical particles in a uniaxial field for various dielectric contrasts: 
(a) $\beta>0, \beta'<0$ or $\ep_1>\ep_2>\ep_1'$, 
(b) $\ep_2>\ep_1=\ep_1'$ in low volume fractions, and 
(c) $\ep_2>\ep_1=\ep_1'$ in high volume fractions.}
\label{figagguf}
\end{figure}

\begin{figure}[h]
\caption{The correction factor of the aggregation time of two spherical 
particles in a uniaxial field plotted against the initial separation. 
The upper panel corresponds to the transverse field case while the lower 
pannel corresponds to the longitudinal field case.}
\label{figratiouf}
\end{figure}

\begin{figure}[h]
\caption{The displacement-time graph for athermal aggregation of two 
spherical particles in a rotating field for various dielectric contrasts, 
rotating field frequencies for (a) low volume fractions and (b) high 
volume fractions.} 
\label{figaggrf}
\end{figure}

\begin{figure}[h]
\caption{The correction factor of the aggregation time of two spherical 
particles in a rotating field plotted against the initial separation for 
two different rotating field frequencies $\omega$, and $\tau_{DID}$ and 
$\tau_{PD}$ are the aggregation times in the DID and PD models 
respectively.} 
\label{figratiorf}
\end{figure}

\newpage
\centerline{\epsfig{file=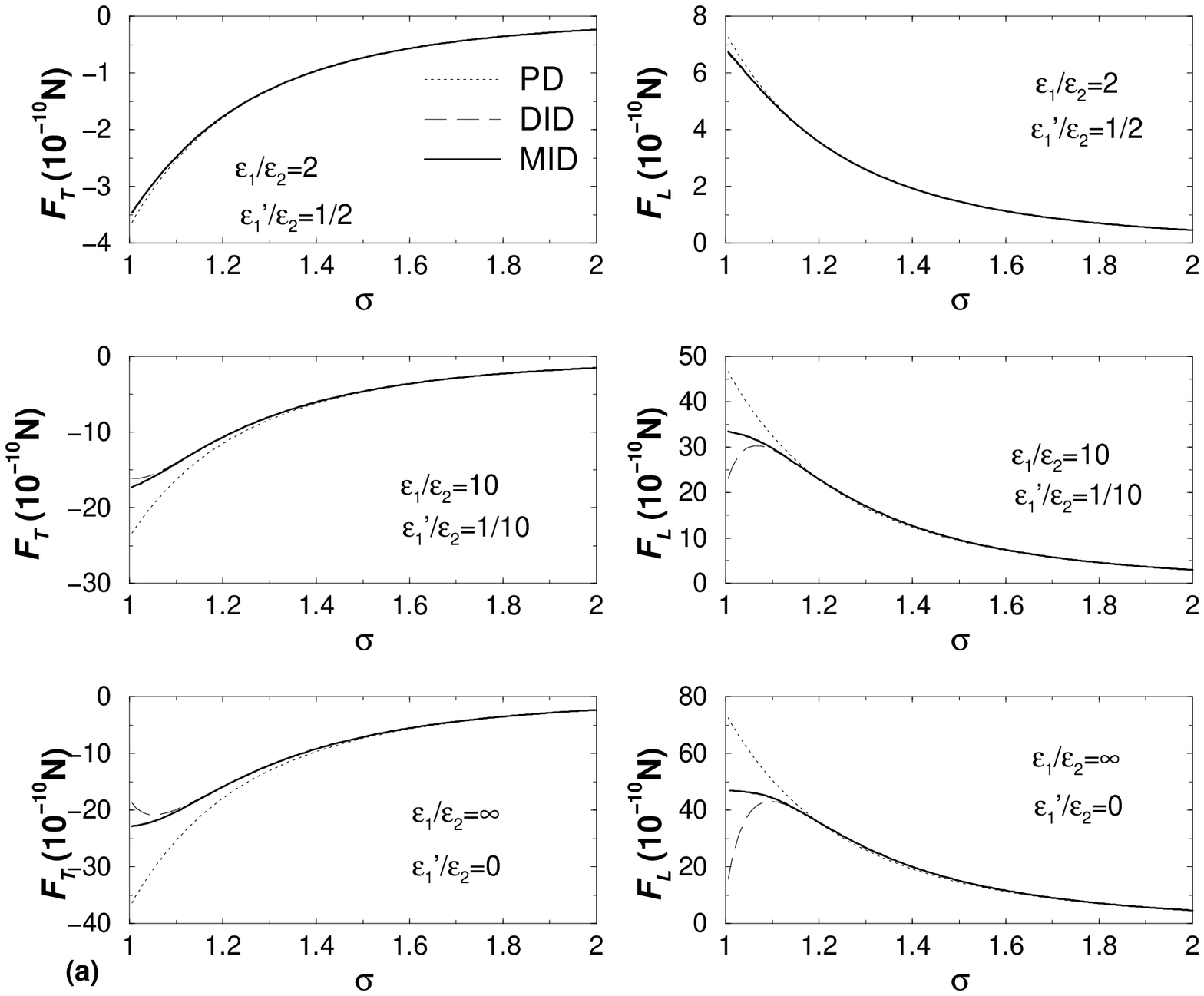,width=\linewidth}}
\centerline{Fig.1(a)/Wong, Sun and Yu}

\newpage
\centerline{\epsfig{file=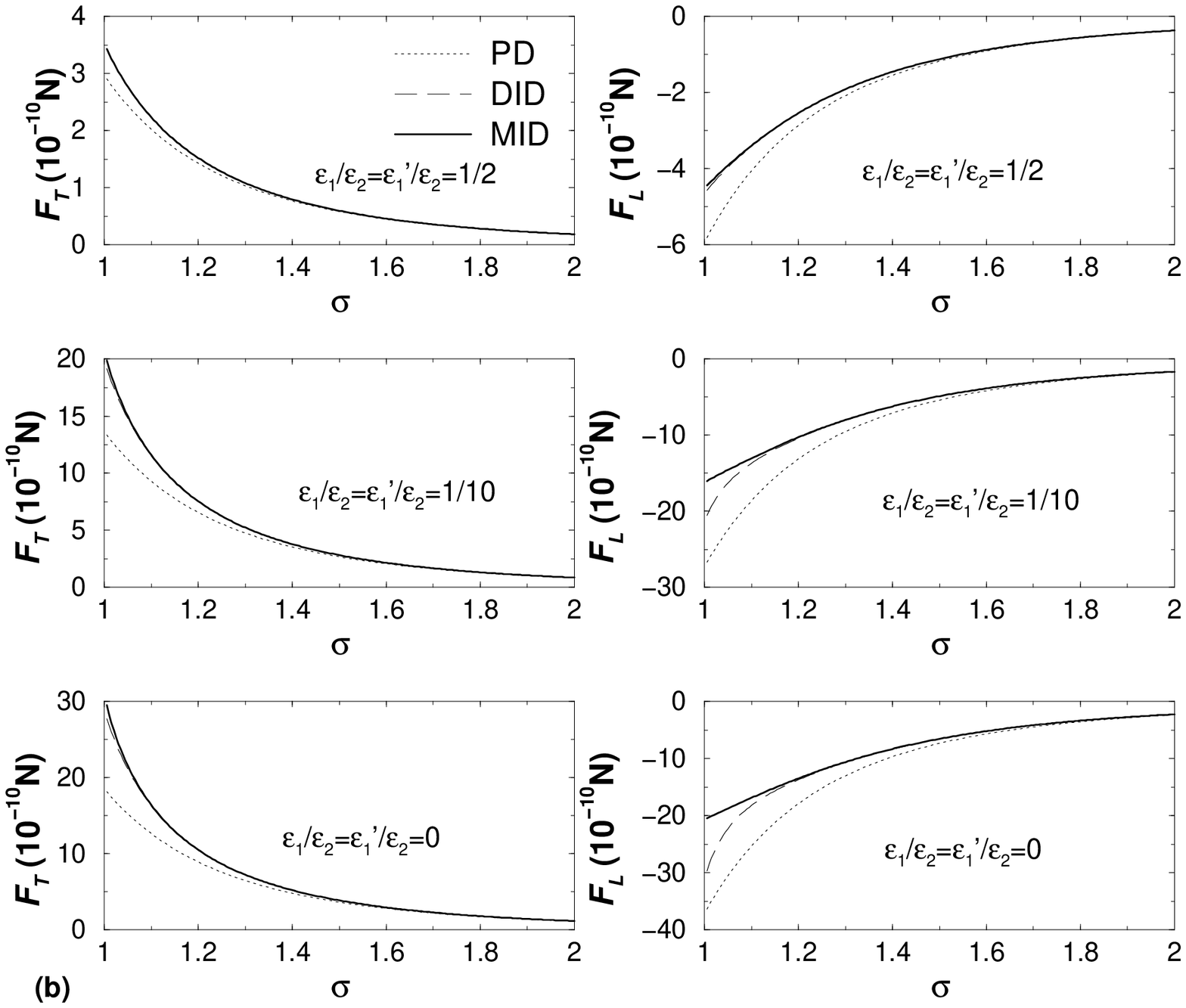,width=\linewidth}}
\centerline{Fig.1(b)/Wong, Sun and Yu}

\newpage
\centerline{\epsfig{file=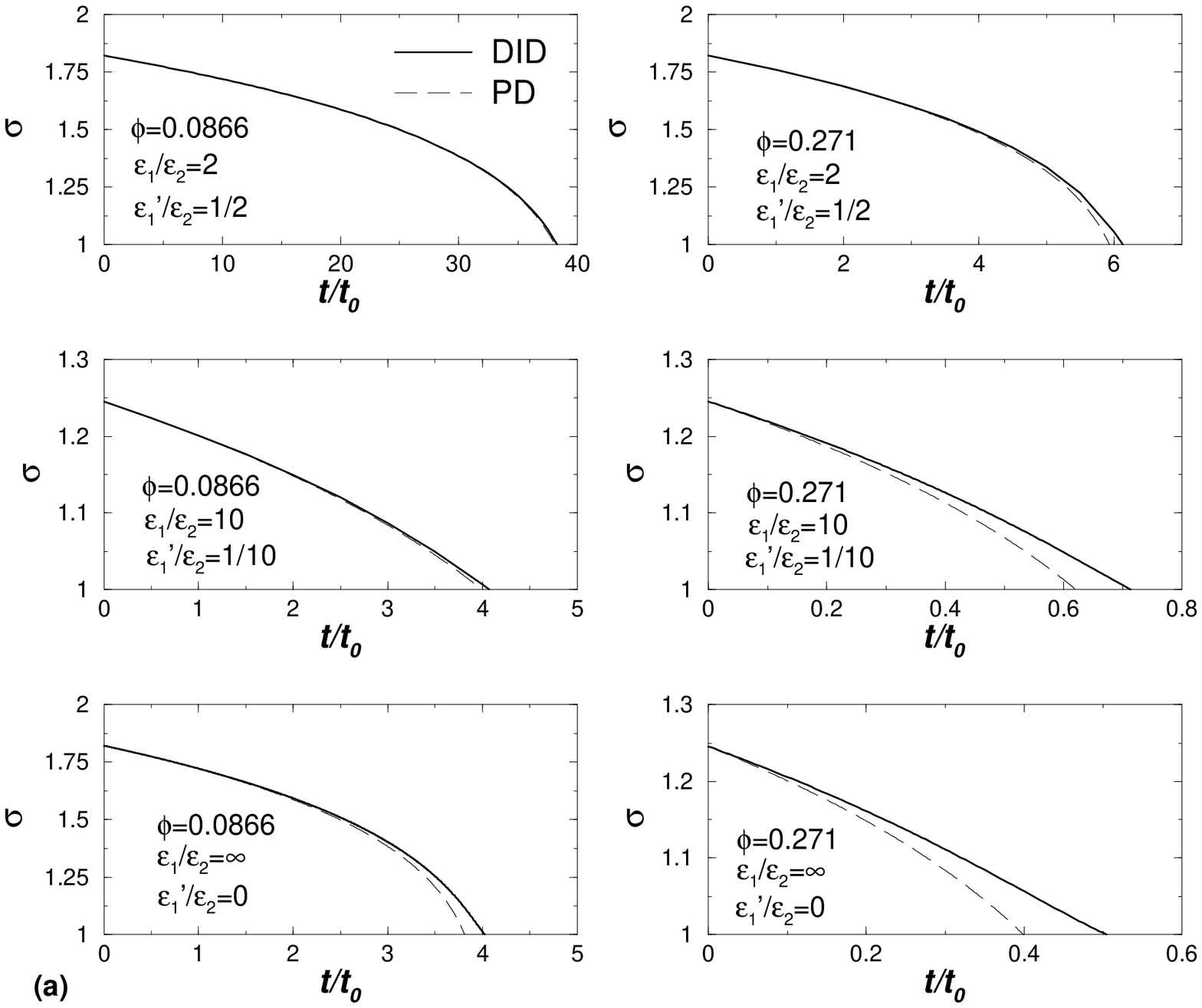,width=\linewidth}}
\centerline{Fig.2(a)/Wong, Sun and Yu}

\newpage
\centerline{\epsfig{file=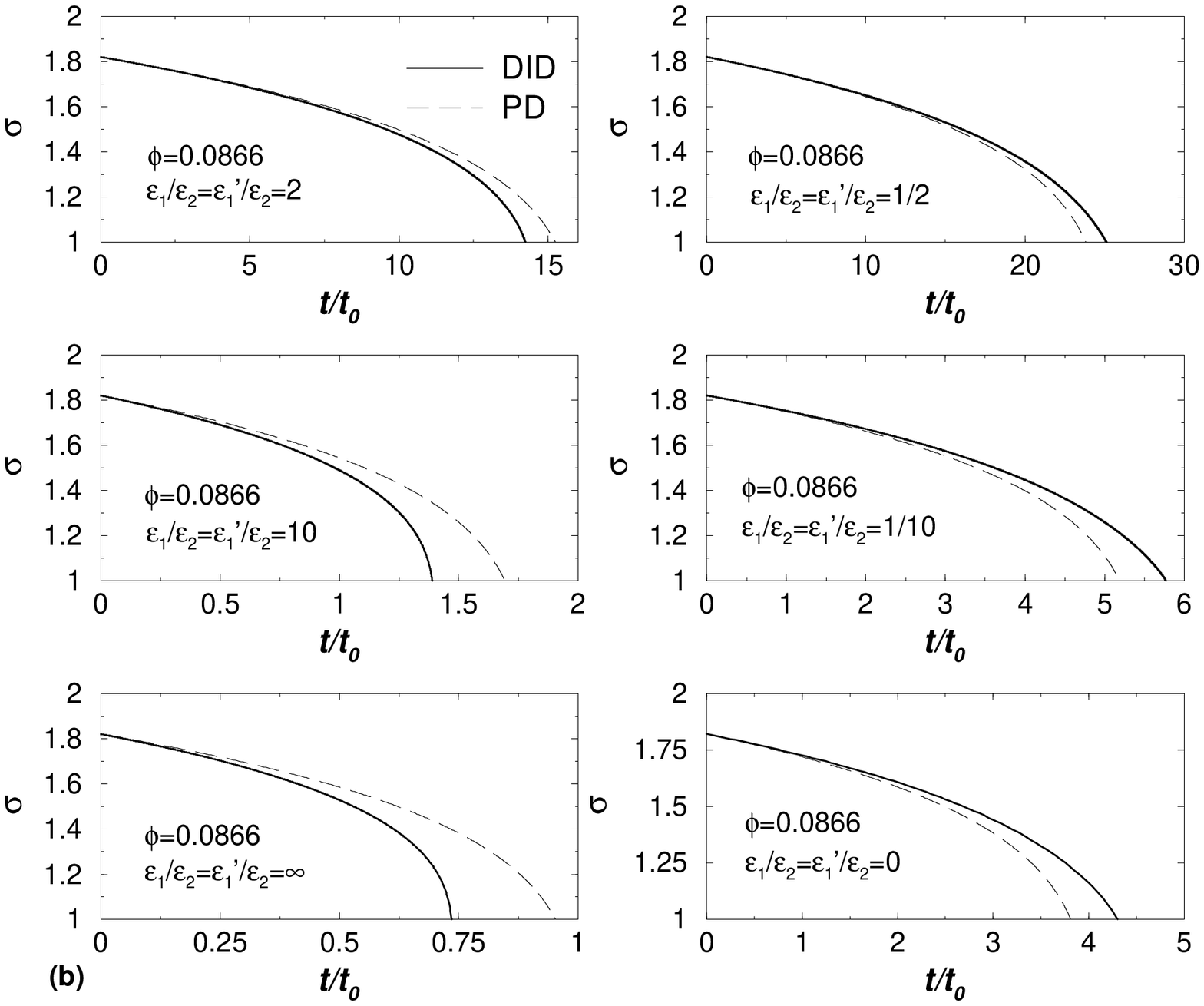,width=\linewidth}}
\centerline{Fig.2(b)/Wong, Sun and Yu}

\newpage
\centerline{\epsfig{file=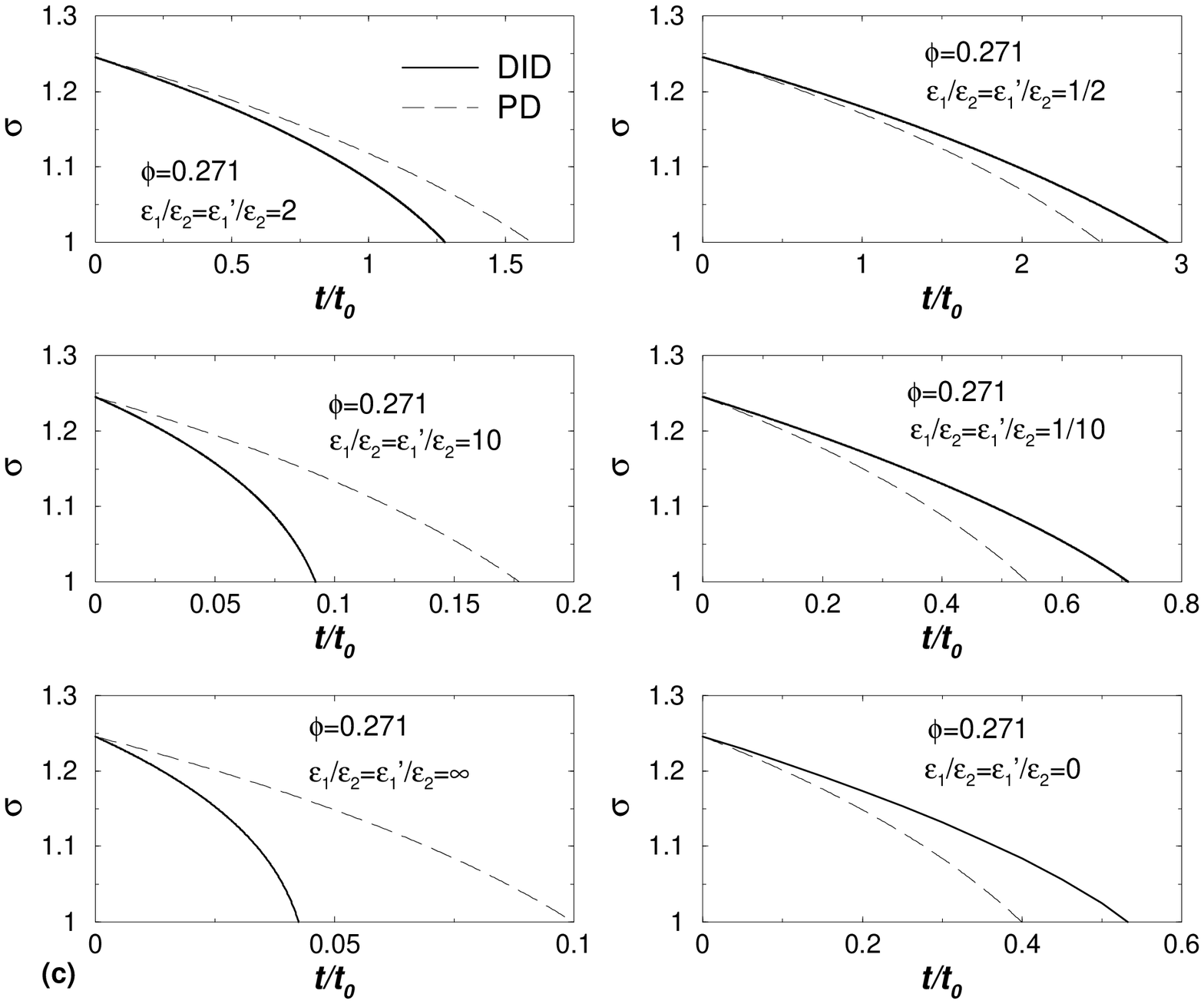,width=\linewidth}}
\centerline{Fig.2(c)/Wong, Sun and Yu}

\newpage
\centerline{\epsfig{file=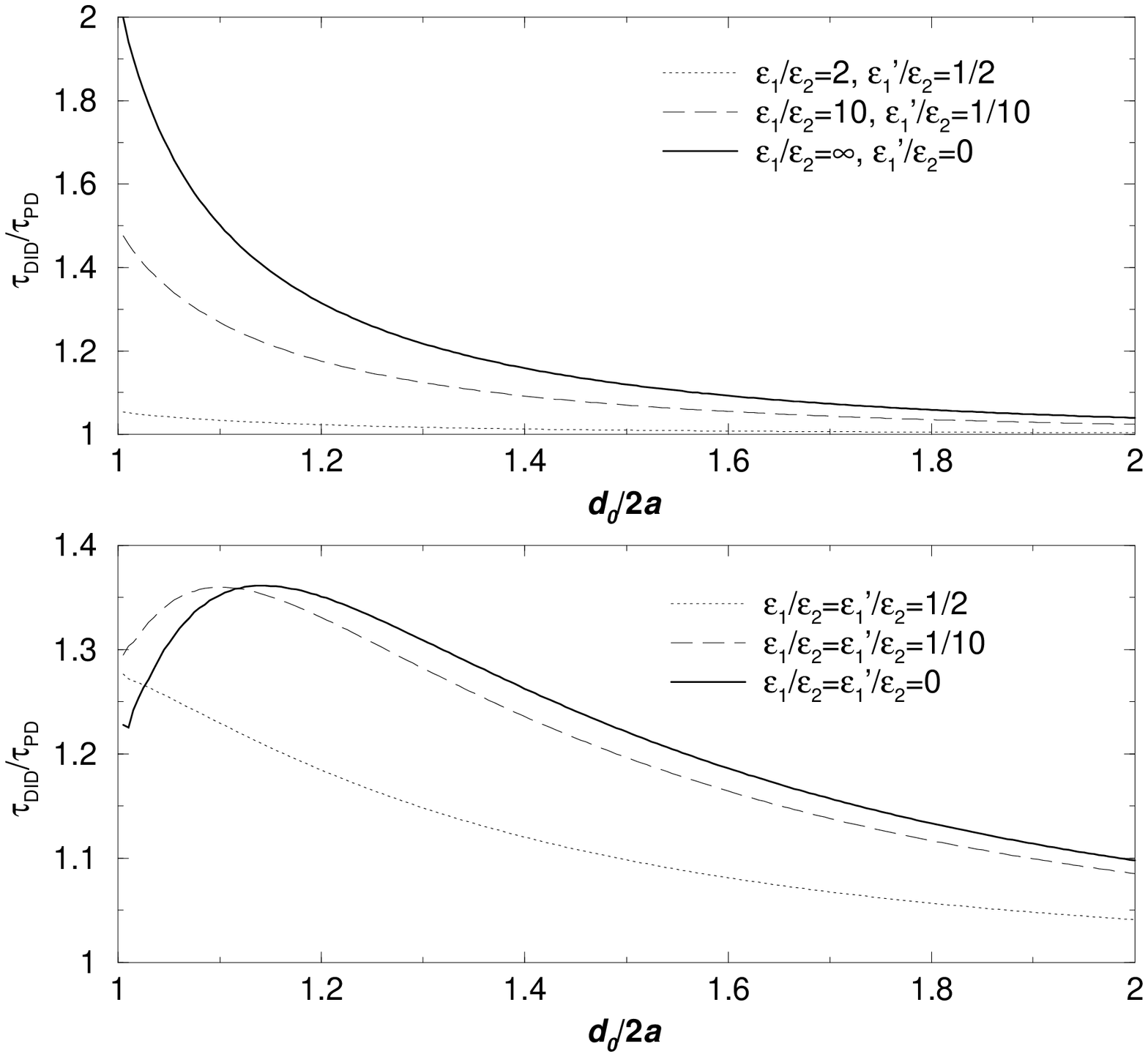,width=\linewidth}}
\centerline{Fig.3/Wong, Sun and Yu}

\newpage
\centerline{\epsfig{file=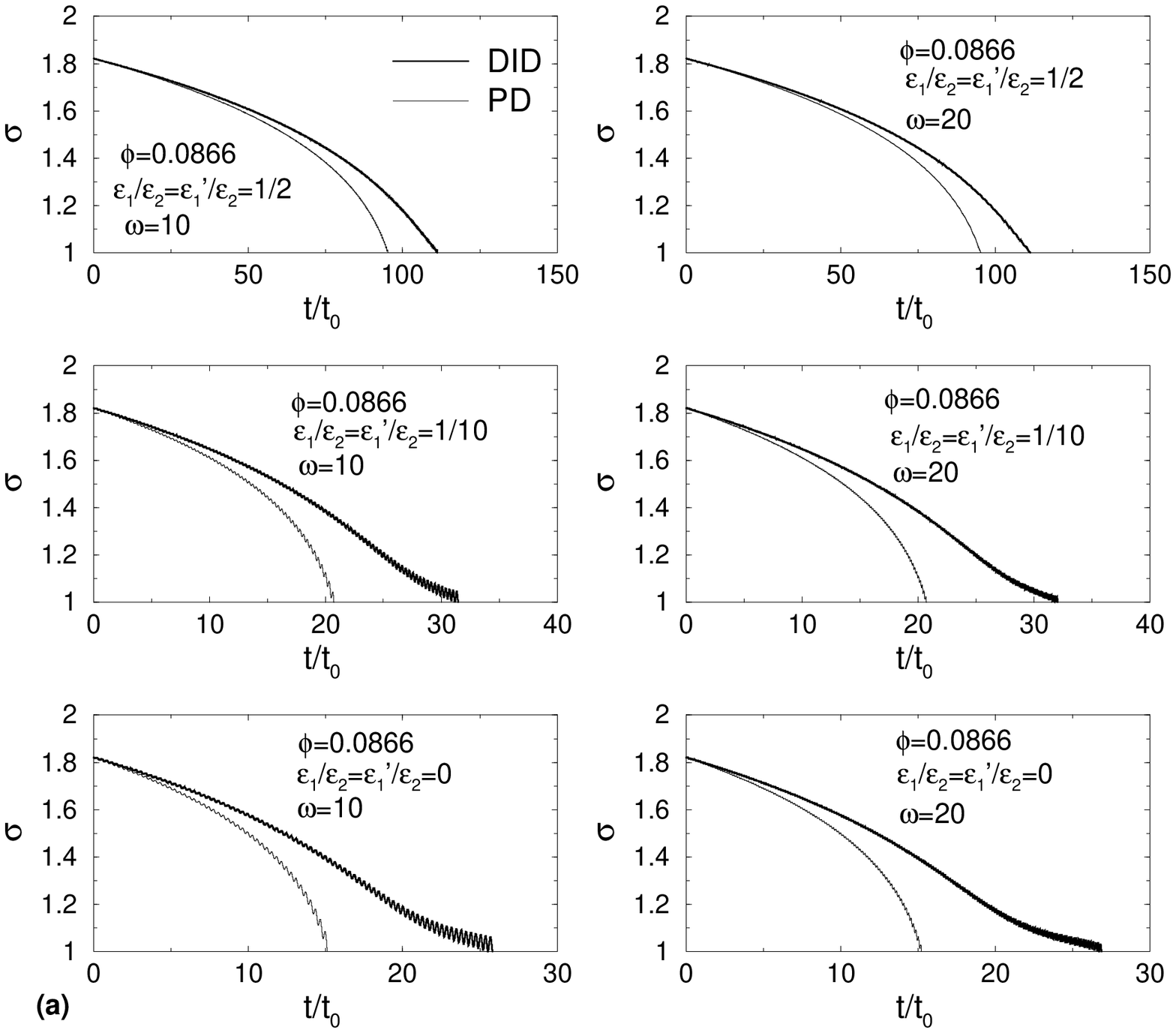,width=\linewidth}}
\centerline{Fig.4(a)/Wong, Sun and Yu}

\newpage
\centerline{\epsfig{file=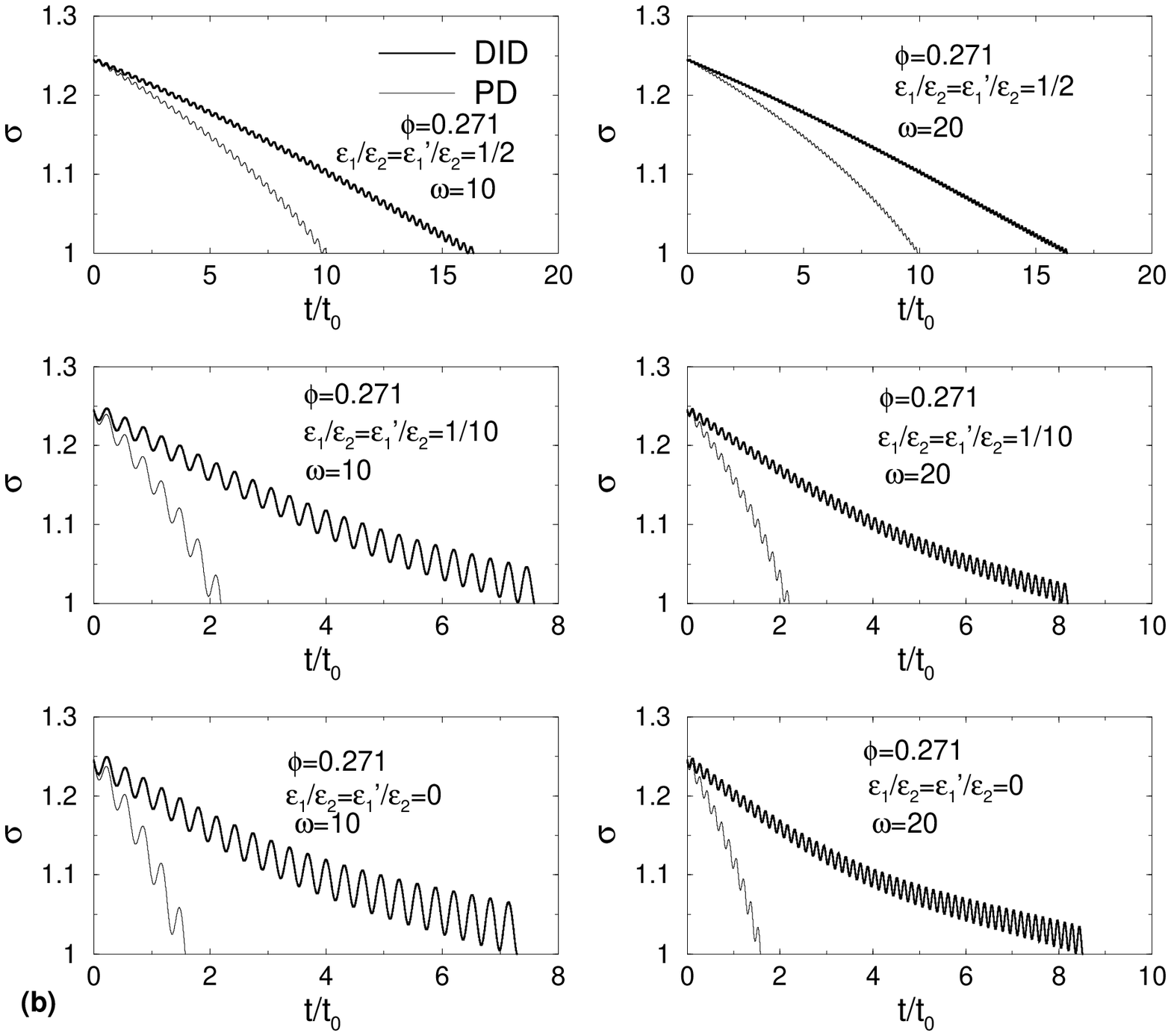,width=\linewidth}}
\centerline{Fig.4(b)/Wong, Sun and Yu}

\newpage
\centerline{\epsfig{file=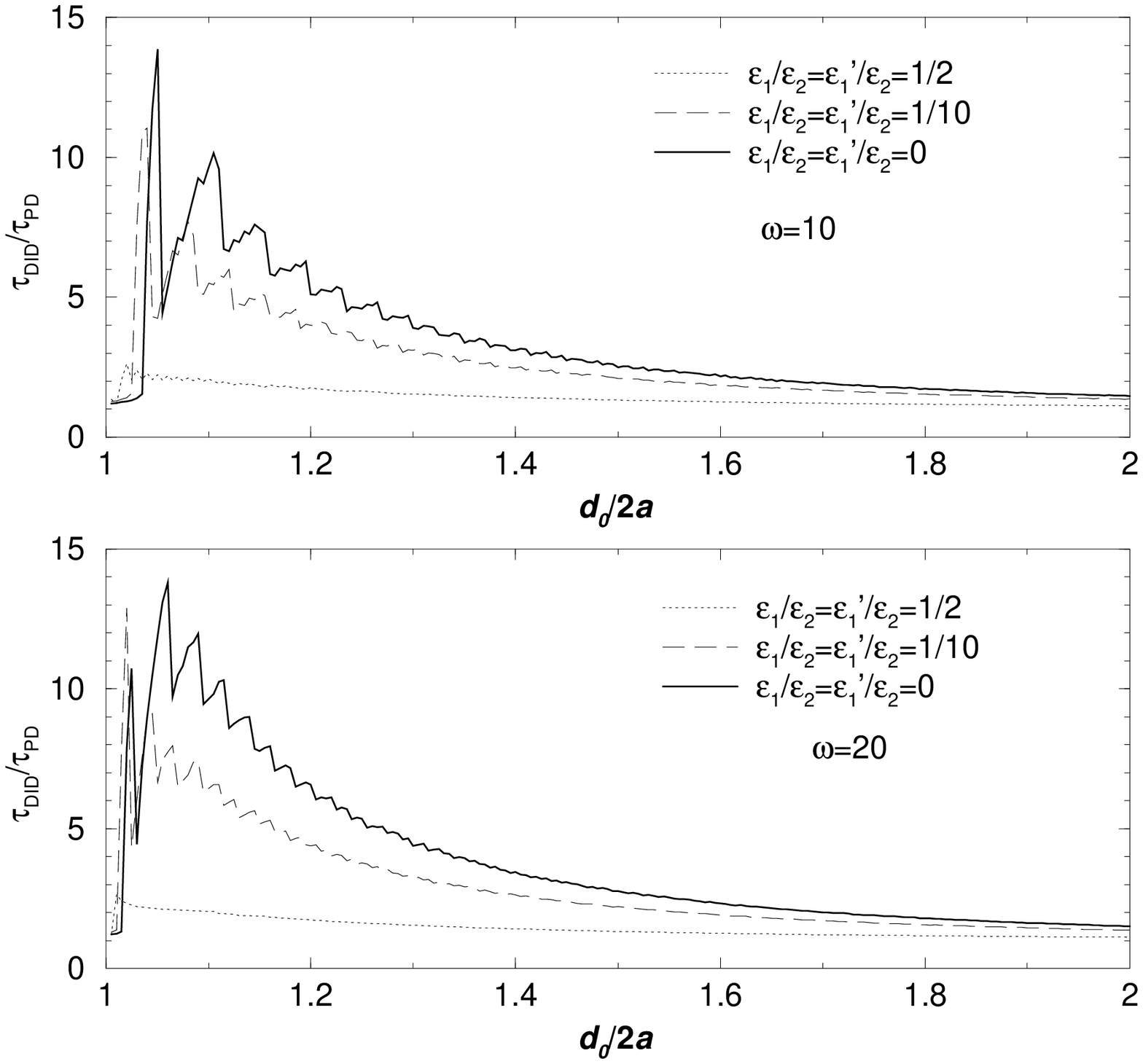,width=\linewidth}}
\centerline{Fig.5/Wong, Sun and Yu}

\end{document}